\documentclass[aps,twocolumn,superscriptaddress]{revtex4-2}

\usepackage{graphicx}
\usepackage{dcolumn}
\usepackage{bm}
\usepackage{hyperref}

\usepackage[caption=false]{subfig}
\usepackage{adjustbox}

\usepackage{amssymb,amsthm}

\theoremstyle{definition}

\usepackage{quantikz}
\usepackage{ulem}

\begin{document}


\title{Universality of a standard
  two-qubit gate by catalytic embedding}

\author{Robin Kaarsgaard}
\email{kaarsgaard@imada.sdu.dk}
\affiliation{%
  Centre for Quantum Mathematics,
  University of Southern Denmark,
  Odense,
  Denmark
}%

\date{\today}

\begin{abstract}
We study the resources required to achieve universal quantum computing
via the gate sets that provide the fundamental instructions from which
quantum algorithms are built. While single-gate universal sets are
known, they rely on precisely tuned irrational rotations, making them
difficult to realize on near-term devices. We find that the
controlled-$V$ gate, an elementary two-qubit interaction directly
implementable on leading hardware, is universal and capable of
simulating standard universal gate sets with minimal
overhead. Specifically, we use catalytic embeddings to develop a
constant-overhead algorithm that simulates standard universal gate
sets, including Clifford$+T$ and Clifford$+$Toffoli. We combine this
simulation algorithm with existing synthesis results to yield exact
and approximate synthesis algorithms for unitaries with and without
number-theoretic restrictions. The results highlight how full quantum
computational power, complete with algorithms for synthesis and
simulation, can emerge from unexpectedly simple ingredients.
\end{abstract}

\maketitle

Quantum computers promise capabilities beyond those of classical
devices~\cite{supremacy}, as exemplified by Shor’s algorithm for
factoring~\cite{shor}, Grover’s search algorithm~\cite{grover}, and
algorithms for Hamiltonian
simulation~\cite{10.1145/3313276.3316366}. To analyse and implement
such algorithms, one requires a universal gate set that provides a
complete basis for quantum computation, just as the $\mathit{NAND}$
gate underpins classical circuit logic. Identifying minimal or
structurally simple universal gate sets is therefore of both
theoretical and practical interest, and also sheds light on the
essential resources that distinguish quantum from classical
computation.

Among the many possible two-qubit interactions, the controlled-$V$
($CV$) gate occupies an interesting position. Previously described as
``semi-classical''~\cite{abdessaied2016complexity}, it has been
extensively studied in the quantum synthesis of classical reversible
circuits through the NCV gate set (see, e.g.,
\cite{sasanian2012realizing, abdessaied2016technology,
  abdessaied2016complexity}), which can be generated entirely from
$CV$. Explicitly,
\begin{equation}
  CV = \begin{pmatrix}
    1 & 0 & 0 & 0 \\
    0 & 1 & 0 & 0 \\
    0 & 0 & \tfrac{1+i}{2} & \tfrac{1-i}{2} \\
    0 & 0 & \tfrac{1-i}{2} & \tfrac{1+i}{2}
  \end{pmatrix}.
\end{equation}
The gate is also known as the controlled-$\sqrt{X}$, since applying it
twice yields the familiar controlled-$X$ ($CX$):
\begin{equation}
  CV^2 = CX =
  \begin{pmatrix}
    1 & 0 & 0 & 0 \\
    0 & 1 & 0 & 0 \\
    0 & 0 & 0 & 1 \\
    0 & 0 & 1 & 0
  \end{pmatrix}.
\end{equation}
Consequently, its implementation is related directly to the $CX$ gate,
as it can be obtained by halving the pulse duration on trapped-ion and
superconducting platforms~\cite{9762485}, making it accessible on a
wide variety of quantum hardware today. It has additionally seen
applications in quantum programming languages~\cite{easyaspi} and
compilation~\cite{10.1145/3656436}. Despite this, the computational
power of $CV$ in isolation has remained unresolved.

Previous work on single-gate universality, beginning with
Deutsch~\cite{deutsch1989quantum},
Barenco~\cite{barenco1995universal}, and Sleator and
Weinfurter~\cite{PhysRevLett.74.4087}, relies on control parameters
corresponding to rotation angles that are irrational multiples of
$2\pi$. This condition ensures that repeated applications generate an
infinite set of single-qubit rotations, leading to dense coverage of
the unitary group. Rational cases, such as $CV$, fall outside
of this framework.

In this work, we show that $CV$ alone is capable of universal quantum
computation, using a generalisation of Aharonov's
technique~\cite{aharonov:toffolihadamard} known as catalytic
embedding~\cite{amyetal:catalytic} coupled with a procedure for
generating the necessary resource states. The resulting algorithms simulate
universal gate sets such as (variants of) Clifford+Toffoli
$\{S,H,CX,CCX\}$~\cite{amy2020number} and Clifford$+T$
$\{H,T,CX\}$, all with constant overhead and a small number of (clean)
auxiliary qubits, assuming only nearest-neighbour connectivity and
access to $CV$. We extend the simulation algorithms to an exact
synthesis algorithm for unitaries with entries in the number ring
$\mathbb{Z}[\tfrac{1}{2},i]$, and an approximate synthesis algorithm
for unitaries with entries in $\mathbb{Z}[\tfrac{1}{2},\omega]$ (with
$\omega$ an eighth root of unity) which is exact in the presence of
a single $\ket{T}$-state. This reinforces the intuition that universal
quantum computing requires no more than the tools of classical circuit
logic supplemented by coherent
superposition~\cite{shi:aharonov,aharonov:toffolihadamard,easyaspi}.

\section{Computational universality and catalytic embeddings}
Universality for quantum computing is commonly used synonymously with
denseness as a subgroup of the unitary group: a gate set $G$ is
(strictly) universal in case there exists $k$ such that for all $n \ge
k$, the group $G_n$ of $n$-qubit circuits with gates taken from $G$ is
dense in $U(2^n)/U(1)$. However, this can be argued to be too strong
to be useful in practice~\cite{aharonov:toffolihadamard}, as it
permits neither the use of auxiliary qubits (which are otherwise
commonly used in
synthesis~\cite{shi:aharonov,PhysRevA.87.032332,cyclotomic}), nor the
use of encodings (which are used in fault-tolerant quantum
computing~\cite{baconetal:encoded}). A caveat to more permissive
notions of universality is that
Solovay-Kitaev~\cite{kitaev1997quantum} does not apply directly, so a
fast rate of approximation is not guaranteed unless the encoding size
is carefully controlled. Accounting for this leads to the notion of a
computationally universal gate
set~\cite[Def.~3]{aharonov:toffolihadamard}: a gate set that can be
used to simulate to within $\epsilon$ error any quantum circuit
which uses $n$ qubits and $t$ gates from a strictly universal gate set
with only polylogarithmic overhead in $\left(n, t,
\tfrac{1}{\epsilon}\right)$.

\begin{figure}
    \subfloat[][]{\label{fig:vdown}
      \begin{adjustbox}{height=4ex}
        \begin{quantikz}
          & \ctrl{1} & \\
          & \gate{V} &
        \end{quantikz}
      \end{adjustbox}
    } \quad
    \subfloat[][]{\label{fig:vup}
      \begin{adjustbox}{height=4ex}
        \begin{quantikz}
          & \gate{V}  &  \\
          & \ctrl{-1} &
        \end{quantikz}
      \end{adjustbox}
    } \quad
    \subfloat[][]{\label{fig:cnot}
      \begin{adjustbox}{height=4ex}
        \begin{quantikz}
          & \ctrl{1} & \\
          & \targ{} &
        \end{quantikz}
        \enspace = \enspace 
        \begin{quantikz}
          & \ctrl{1} & \ctrl{1} & \\
          & \gate{V} & \gate{V} &
        \end{quantikz}
      \end{adjustbox}
    }
    
    \subfloat[][]{\label{fig:cvinv}
      \begin{adjustbox}{height=4ex}
        \begin{quantikz}
          & \ctrl{1} & \\
          & \gate{V^\dagger} &
        \end{quantikz}
        \enspace = \enspace 
        \begin{quantikz}
          & \ctrl{1} & \ctrl{1} & \\
          & \targ{}  & \gate{V} &
        \end{quantikz}
      \end{adjustbox}
    } \enspace
    \subfloat[][]{\label{fig:swap}
      \begin{adjustbox}{height=4ex}
        \begin{quantikz}
          & \swap{1} & \\
          & \targX{} &
        \end{quantikz}
        \enspace = \enspace 
        \begin{quantikz}
          & \ctrl{1} & \targ{}   & \ctrl{1} & \\
          & \targ{V} & \ctrl{-1} & \targ{}  &
        \end{quantikz}
      \end{adjustbox}
    }
    
    \subfloat[][]{\label{fig:toff}
      \begin{adjustbox}{height=6ex}
        \begin{quantikz}
          & \ctrl{1} & \\
          & \ctrl{1} & \\
          & \targ{} &
        \end{quantikz}
        \enspace = \enspace 
        \begin{quantikz}
          &          & \ctrl{1} &          & \ctrl{1} & \ctrl{2}    & \\
          & \ctrl{1} & \targ{}  & \ctrl{1} & \targ{}  &             & \\
          & \gate{V} &              & \gate{V^\dagger} &    & \gate{V} &
        \end{quantikz}
      \end{adjustbox}
  }
  \caption{We assume that both nearest-neighbour configurations of the
    $CV$ gate shown in (a) and (b) are available, and
    use these to derive (c) $CX$, (d)
    $CV^\dagger$, (e) $\mathit{SWAP}$, and (f) the
    Toffoli gate by the Sleator-Weinfurter
    construction~\cite{PhysRevA.52.3457}.}
  \label{fig:classical}
\end{figure}

Catalytic embedding~\cite{amyetal:catalytic} is a recent technique for
circuit encoding and synthesis~\cite{cyclotomic} that has proven
fruitful in establishing universality
results~\cite{PhysRevLett.133.050601,
  kissinger2024catalysing}. Catalytic embeddings rely on certain
resource states, catalysts, passed as states of an auxiliary system to
perform computation: as the name suggests, these catalysts (unlike
magic states) are not consumed during computation, but are returned
unchanged for later reuse. In more detail, when $U$ is a unitary on
$\mathcal{H}$, a catalytic embedding of $U$ consists of an auxiliary
system $\mathcal{K}$, a state $\ket{\chi}$ on $\mathcal{K}$ (the
catalyst), and a unitary $\Gamma_U$ on $\mathcal{H} \otimes
\mathcal{K}$ satisfying
\begin{equation}\label{eq:catalytic}
  \Gamma_U \ket{\psi}\ket{\chi} = (U \ket{\psi})\ket{\chi}
\end{equation}
for all states $\ket\psi$ on $\mathcal{H}$. One can think of
$\Gamma_U$ as an encoding of $U$ which can be decoded by passing the
catalyst $\ket{\chi}$ to the auxiliary system.  A trivial example of a
catalytic embedding is the simulation of a gate $U$ using a
controlled-$U$ gate with $\ket1$ as catalyst. A more interesting
example is the encoding of (complex) unitaries as orthogonal
matrices~\cite[Def.~1]{aharonov:toffolihadamard}: given unitary $U$ on
some $\mathcal{H}$, this encoding constructs $\tilde{U}$ accepting an
auxiliary qubit such that
\begin{align}
  \tilde{U}\ket\psi\ket0 & = \Re(U)\ket\psi\ket0 +
  \Im(U)\ket\psi\ket1 \label{ah1} \\
  \tilde{U}\ket\psi\ket1 & = \Re(U)\ket\psi\ket1 -
  \Im(U)\ket\psi\ket0 \label{ah2}
\end{align}
where $\Re(U)$ and $\Im(U)$ denote respectively the entrywise real and
imaginary components of $U$. This is a catalytic embedding with
auxiliary system $\mathbb{C}^2$ and catalyst $\ket\downarrow =
\tfrac{1}{\sqrt{2}}(\ket0 - i\ket1)$, as it follows from
\eqref{ah1} and \eqref{ah2} above that
\begin{equation}
  \tilde{U}\ket\psi\ket\downarrow = (U\ket\psi)\ket\downarrow
\end{equation}
for all states $\ket\psi$ on $\mathcal{H}$. Block-encodings (central
to the quantum singular value
transformation~\cite{10.1145/3313276.3316366}) are likewise canonical
examples of catalytic embeddings---in fact, every catalytic embedding
is unitarily similar to a
block-encoding~\cite[Prop.~2]{amyetal:catalytic}.

\section{Results}
Before we turn to the simulation algorithm, we briefly explore what
can be expressed without it. Using standard constructions from the
literature~\cite{PhysRevA.52.3457}, we see that the classical
reversible $\mathit{SWAP}$, $CX$, and $CCX$ (Toffoli) gates can be
expressed using $CV$ alone, as shown in Fig.~\ref{fig:classical}. In
fact, the construction of the Toffoli gate generalises to any number
of control wires~\cite[Lemma 7.1]{PhysRevA.52.3457}, though the gates
shown in Fig.~\ref{fig:classical} suffice to do universal reversible
classical computation in the presence of a single auxiliary qubit. It
follows that the $CV$ gate alone is (exactly) universal for classical
reversible computation.

It follows from the implementation of the $\mathit{SWAP}$ gate that
the $CV$ gate is as expressive on architectures with nearest-neighbour
connectivity as it is on those with all-to-all connectivity, since
multi-qubit gates can be applied in arbitrary (non-neighbouring)
configurations if necessary by conjugating with an appropriate
permutation of wires. However, applying multi-qubit gates in
non-neighbouring configurations in this way incurs a cost of $12$ $CV$
gates for each wire that the control must cross, so preference in
circuit routing should still be given to neighbouring gate
configurations.

\begin{figure}
  \subfloat[][]{ \label{fig:enc}
    $\begin{aligned}
      \mathcal{E}\left(\begin{quantikz}
        & \gate{V} &
      \end{quantikz}\right)
      \enspace & = \enspace 
      \begin{quantikz}
        & \gate{V} & \\
        \lstick{$\alpha$} & \ctrl{-1} &
      \end{quantikz} \\
      \mathcal{E}\left(
      \begin{quantikz}
        & \gate{S} & 
      \end{quantikz}\right)
      \enspace & = \enspace 
      \begin{quantikz}
        & \ctrl{1} & \\
        \lstick{$\beta$} & \gate{V}  &
      \end{quantikz} \\
      \mathcal{E}\left(
      \begin{quantikz}
        & \gate{T} &
      \end{quantikz}\right)
      \enspace & = \enspace 
      \begin{quantikz}
        & \ctrl{1} & \ctrl{1} & \\
        \lstick{$\gamma$} & \targ{}   & \gate{S}  &
      \end{quantikz}
    \end{aligned}$
  }
  
  \subfloat[][]{\label{fig:had}
    \begin{quantikz}
      & \gate{H} & 
    \end{quantikz}
    \enspace = \enspace 
    \begin{quantikz}
      & \gate{V} & \gate{S} & \gate{V} &
    \end{quantikz}
  }
  
  \subfloat[][]{\label{fig:cs}
    \begin{quantikz}
      & \ctrl{1} & \\
      & \gate{S} &
    \end{quantikz}
    \enspace = \enspace 
    \begin{quantikz}
      &          & \ctrl{1} &          & \\
      & \gate{H} & \gate{V} & \gate{H} &
    \end{quantikz}
  }

  \subfloat[][]{\label{fig:ch}
      \begin{quantikz}
      & \ctrl{1} &\\
      & \gate{H} &
    \end{quantikz}
    =
    \begin{quantikz}
      & \ctrl{1} & \ctrl{1} & \ctrl{1} & \gate{T^\dagger} & \\
      & \gate{V} & \gate{S} & \gate{V} &                &
    \end{quantikz}
  }

  \caption{The encoding of the gates $V$, $S$, and $T$ is defined by
    the function $\mathcal{E}(-)$ is shown in (a), relying on three
    named auxiliary qubits $\alpha$, $\beta$, and $\gamma$.  The
    encoding of $V$ and $S$ in turn allows one to derive (b) the
    Hadamard gate (up to a global phase) and (c) the controlled $S$
    gate, and (d) the controlled Hadamard gate using standard circuit identities.  The encoding of $T$
    relies on the encoding of $V$ and $S$ (and subsequent derivations
    of $H$ and $CS$).}
  \label{fig:encoding}
\end{figure}

\subsection{Simulation}\label{sub:simulation}
The simulation algorithm takes the form of a catalytic embedding with
an auxiliary system consisting of three qubits, which we name
$\alpha$, $\beta$, and $\gamma$. The idea is to perform three
catalytic embeddings in succession: the first to simulate $V$ gates,
the second to simulate $S$ gates, while the third uses the embedding
from \cite{cyclotomic,kissinger2024catalysing} to simulate a $T$ gate
from the available $CS$ gate. The resulting encodings of the $V$, $S$,
and $T$ gates are shown in Fig.~\ref{fig:encoding}. It can be verified
by direct computation that
\begin{align}
  \mathcal{E}(V)\ket\psi\ket1 & = (V\ket\psi)\ket1 \label{eq:catv} \\
  \mathcal{E}(S)\ket\psi\ket- & = (S\ket\psi)\ket- \label{eq:cats} \\
  \mathcal{E}(T)\ket\psi\ket{T} & = (T\ket\psi)\ket{T} \label{eq:catt}
\end{align}
for all states $\ket\psi$, where $\ket{T} =
\tfrac{1}{\sqrt{2}}\left(\ket0 + \tfrac{1+i}{\sqrt{2}}\ket1\right)$,
showing that each of the three encodings satisfy condition
\eqref{eq:catalytic} for a catalytic embedding.

The simulation algorithm then proceeds as follows, given a circuit $C$
formed using either of the strictly universal Clifford+Toffoli or
Clifford$+T$:
\begin{enumerate}
\item Allocate three auxiliary qubits $\alpha$, $\beta$, and $\gamma$.
\item Apply the equations in Fig.~\ref{fig:classical} and
  Fig.~\ref{fig:had} until all $CX$, $CCX$, and $H$ gates have
  been decomposed into $CV$, $V$, and $S$ gates.
\item Apply the encodings in Fig.~\ref{fig:enc}
  until a circuit consisting of only $CV$ gates is obtained.
\item If any auxiliary qubits are entirely unused (meaning no $CV$ gate
  is applied as either control or target to the qubit), remove them.
\item Generate the resource states $\ket1$, $\ket-$, and $\ket{T}$
  (see \ref{sub:injection}) and pass them to $\alpha$, $\beta$, and
  $\gamma$ respectively.
\end{enumerate}
Notice that the auxiliary $\ket{T}$-state is only necessary when
simulating Clifford+$T$ circuits (since the auxiliary qubit $\gamma$
is only used when simulating $T$ gates); for Clifford+Toffoli, $\ket1$
and $\ket-$ suffice.

Applying this algorithm results in a new circuit $\Gamma_C$ (see
Fig.~\ref{fig:example} for an example of encoding the $CS$ gate). By
\eqref{eq:catv}--\eqref{eq:catt}, the resulting circuit satisfies
\begin{equation}
  \Gamma_C \ket\psi \ket1\ket-\ket{T} = (C \ket\psi) \ket1\ket-\ket{T}
\end{equation}
for all states $\ket\psi$, verifying correctness of the algorithm. As
such, $\Gamma_C$ simulates $C$ with constant additive overhead in
qubit count of up to $3$ additional qubits, and constant
multiplicative overhead of up to $9$ native gates per gate in the
circuit (see also Table \ref{tab:cost}).

\begin{figure}
  \subfloat[][]{\label{fig:csex}
\begin{quantikz}
                    &          &           &          & \ctrl{1} &          &           &          & \\
                    & \gate{V} & \ctrl{2} & \gate{V} & \gate{V} & \gate{V} & \ctrl{2} & \gate{V} & \\
  \lstick{$\alpha$} & \ctrl{-1} &           & \ctrl{-1} &          & \ctrl{-1} &           & \ctrl{-1} & \\
  \lstick{$\beta$}  &          & \gate{V}  &          &          &          & \gate{V}  &          &
\end{quantikz}}
\quad
  \subfloat[][]{\label{fig:minusinj}
      \begin{quantikz}
        \lstick{\ket0} & \gate{V} & \targ{}   & \ctrl{1} & \meter{} \\
        \lstick{\ket0} & \gate{V} & \ctrl{-1} & \targ{}  &
      \end{quantikz}
  }
\caption{Circuits related to the encoding: (a) The $CV$ circuit
  encoding a $CS$ gate by decomposing it into $V$, $S$, and $CV$ gates
  using the equations in Fig.~\ref{fig:encoding} and encoding the $V$
  and $S$ gates. The auxiliary qubit $\gamma$ is unused and has been
  omitted. (b) The circuit producing a $\ket-$ state when the
  measurement outcome is $1$ (corresponding to the eigenstate
  $\ket0$), which occurs with probability $\tfrac{1}{2}$.}
\label{fig:example}
\end{figure}

Clifford+Toffoli and Clifford+$T$ are not the only universal gate sets
that can be simulated this way. Extending the second step in the
algorithm with the equations in Fig.~\ref{fig:cs} and
Fig.~\ref{fig:ch} allows the simulation of circuits formed using the
universal Clifford+$CS$, Clifford+$CH$, and
$\{H,CS\}$~\cite{kitaev1997quantum} gate sets.

\subsection{Resource state injection}\label{sub:injection}
The crux of the simulation is the availability of the resource states
$\ket1$, $\ket-$, and $\ket{T}$. While only a single state of each is
required, of these only $\ket1$ can be reasonably assumed to be
immediately available (we discuss later how this assumption can be
relaxed). We address this in two stages, corresponding to the two
remaining catalysts $\ket-$ and $\ket{T}$.

First, the $\ket-$ state can be prepared by executing the circuit
shown in Fig.~\ref{fig:minusinj}, which may be implemented using only
computational basis states and $CV$ gates using \eqref{eq:catv} and
the implementations of classical gates in Fig.~\ref{fig:classical}:
when the measurement outcome is $1$, which occurs with probability
$\tfrac{1}{2}$, the second qubit is in the state $\ket-$ (up to a
global phase). This constitutes an $O(1)$ preprocessing step
to the simulation algorithm.

To approximate a $\ket{T}$ state, once a $\ket-$ state has been
produced as above, it is passed along with $\ket1$ to the simulation
algorithm applied to a Clifford+Toffoli circuit that prepares a
$\ket+$ state and applies a $\pi/4$ $Z$-rotation using the
single-qubit rotation algorithm of \cite{hindlycke2024single}. By the
catalytic property, we can recover the $\ket-$ state used during this
simulation, and pass it along with the $\ket{T}$ state produced during
the above procedure to be used for the simulation of Clifford$+T$
circuits. Following \cite{amyetal:catalytic}, approximating the
$\ket{T}$ catalyst up to $\epsilon$ yields an approximate simulation
of any Clifford+$T$ circuit with error $\epsilon$.  Alternatively, it
was argued in \cite{kissinger2024catalysing} that expectations of
observables of Clifford+$T$ circuits encoded this way can be computed
by replacing the $\ket{T}$-state used as catalyst by a carefully
prepared cocktail of stabiliser states.

\subsection{Synthesis}
An immediate application of the simulation algorithm is to combine it
with existing synthesis algorithms in order to synthesise circuits
from unitaries directly. For Clifford+Toffoli specifically, since the
resource state injection protocol produces an exact $\ket-$ state, we
can synthesise a $CV$ circuit exactly from a unitary $U$ with entries in
the number ring $\mathbb{Z}[\tfrac{1}{2},i]$ as follows:
\begin{enumerate}
\item Decompose $U$ into generators $i_{[a]}$, $X_{[a,b]}$, $\omega
  H_{[a,b]}$ using \cite[Algorithm~2.14]{bianselinger}.
\item Synthesise unitary generators into a Clifford+Toffoli circuit
  using Gray codes with at most one clean auxiliary qubit via
  \cite[Proposition~4.6]{amyglaudellross:numbertheoretic}.
\item Apply the simulation algorithm of \ref{sub:simulation} to the
  resulting Clifford+Toffoli circuit.
\end{enumerate}
Since the circuit synthesis algorithm requires up to one auxiliary
qubit, and the simulation algorithm requires a further two in the case
of Clifford+Toffoli, this yields a synthesised circuit for $U$ with
three auxiliary qubits.

The same general approach can be used to synthesise a $CV$ circuit
from a unitary with entries in the ring
$\mathbb{Z}[\tfrac{1}{2},\omega]$ by means of an algorithm that
synthesises to a Clifford$+T$
circuit~\cite{PhysRevA.87.032332,kliuchnikov2013synthesis}, though
with one caveat: since the state injection protocol only realises the
$\ket{T}$ state with imperfect fidelity, the resulting synthesis is
only approximate, subject to the simulation error described in
\ref{sub:injection}. However, the simulation error can be eliminated
by the presence of a single exact $\ket{T}$ state. In fact, this
approach can be extended to synthesise arbitrary unitaries, using the
method in \cite{kliuchnikov2013synthesis} to round off a unitary to
one with entries in $\mathbb{Z}[\tfrac{1}{2},\omega]$ within a given
error.

\section{Implementation}
The $CV$ gate can be realised by pulse-engineering on both
superconducting and trapped-ion devices~\cite{9762485}, requiring no
specialised protocols beyond those already used for $CX$. By contrast,
protocols for realising the universal
Barenco~\cite{PhysRevA.97.032310} and
Deutsch~\cite{PhysRevApplied.9.051001,PhysRevResearch.6.023231} gates
have only fairly recently begun to emerge, primarily in the context of
neutral-atom architectures.

The assumption that both computational basis states can be produced on
demand can be bypassed. On architectures where only ground states
$\ket0$ are readily prepared, the simulation can be adapted by instead
implementing a \textit{negatively}-controlled-$V$ gate. This allows
passing $\ket0$ to the control qubit of this gate to implement a $V$
gate, and, in turn, allows the implementation of an $X$ gate by $V^2$,
making available both $\ket1$ states and the (positively)
controlled-$V$ gate by conjugating negatively-controlled-$V$ on the
control qubit by $X$. The simulation then proceeds as presented.

\section{Discussion and outlook}
The gate set $\{CV,N(\pi/2^n)\}$, where
\begin{equation}
N(\theta) = \frac{1}{2}\begin{pmatrix}
  1+e^{i \theta}   & 1 - e^{i \theta} \\
  1 - e^{i \theta} & 1+e^{i \theta}
\end{pmatrix}
\end{equation}
is the Negator gate~\cite{de2013negator} generalising the $V$ gate in
that $N(\pi/k)^k = X$, was shown in \cite{deVos} to realise all
$n$-qubit permutations without auxiliary qubits; similarly, the
classically universal NCV gate set $\{X,CX,CV,CV^\dagger\}$ has been
widely studied in the quantum synthesis of reversible classical
circuits~\cite{sasanian2012realizing, abdessaied2016technology,
  abdessaied2016complexity}. A natural question is whether their
expressive power extends to full universality, which we show here to
be the case. Similarly, the gate
\begin{equation} \small
S_U^{(\tau)} = \begin{pmatrix}
  1 & 0 & 0 & 0 \\
  0 & 1 & 0 & 0 \\
  0 & 0 & e^{i(\pi/4)} \cos(\pi \tau/2) & e^{-i(\pi/4)} \sin(\pi \tau/2) \\
  0 & 0 & e^{-i(\pi/4)} \sin(\pi \tau/2) & e^{i(\pi/4)} \cos(\pi \tau/2)
\end{pmatrix}
\end{equation}
considered by Sleator and Weinfurter~\cite{PhysRevLett.74.4087} is
precisely the $CV$ gate for $\tau=1/2$~\footnote{There is an
unfortunate printing error in \cite{PhysRevLett.74.4087} describing
$S_U^{(1)}$ rather than $S_U^{(1/2)}$ as the $CV$ gate.}, yet their
derivation of universality relies on choosing an irrational $\tau$
(similar to \cite{barenco1995universal,deutsch1989quantum}). A direct
consequence of this work is that the rational $\tau = 1/2$ suffices.

This work opens the possibility of performing universal fault-tolerant
quantum computing solely by means of a fault-tolerant $CV$
gate. Similarly, just as $T$ gates are considered the central
computational resource in the context of Clifford+$T$ circuits, one
can study minimal $CV$-count of algorithms as a means to quantify the
quantum computational resources necessary to realise them.

\textit{Acknowledgements}---I thank Robert Booth, Chris Heunen,
Nicolas Heurtel, Louis Lemonnier, (Neil) Julien Ross, Alexis De Vos,
and John van de Wetering for their valuable comments, suggestions, and
discussions related to this letter. This research was supported by
Sapere Aude: DFF-Research Leader grant 5251-00024B.

\bibliography{../bibliography}

\begin{table}
  \vspace{\baselineskip}
  \begin{tabular}{l c c c}
    \textbf{Gate} & \phantom{XXX} & \textbf{Qubits} & \textbf{Gates} \\
    \hline
    $CX$ && $0$ & $2$ \\
    $\mathit{SWAP}$ && $0$ & $6$ \\
    $CCX$ && $0$ & $9$ \\
    $V$ && $1$ & $1$ \\
    $X$ && $1$ & $2$ \\
    $Z$ && $1$ & $2$ \\
    $S$ && $1$ & $1$ \\
    $H$ && $2$ & $3$ \\
    $CS$ && $2$ & $7$ \\
    $T$ && $3$ & $9$
  \end{tabular}
\caption{Some common gates and their simulation cost in terms of
  auxiliary qubits and number of $CV$ gates. The gate cost
  of a circuit is the sum of the costs of its constituent gates, while
  the qubit cost depends on the number of unused auxiliary qubits: the
  Clifford+Toffoli gate set $\{S,H,CX,CCX\}$ requires at most $2$
  auxiliary qubits, while Clifford$+T$ $\{H,T,CX\}$ requires at most
  $3$.}
\label{tab:cost}
\end{table}

\appendix

\section{Omitted proofs}
We verify here the correctness of the three catalytic embeddings and the
resource state injection procedure.

\subsection{Catalytic embeddings}
That $\mathcal{E}(V)\ket\psi\ket1 = (V \ket\psi)\ket1$
for all $\ket\psi$ follows by
\begin{align*}
  \mathcal{E}(V)\ket\psi\ket1
  & = (\textit{SWAP} \cdot CV \cdot \textit{SWAP})\ket\psi\ket1 \\
  & = (\textit{SWAP} \cdot CV)\ket1\ket\psi \\
  & = \textit{SWAP} (\ket1(V\ket\psi)) \\
  & = (V\ket\psi)\ket1
\end{align*}
To see that $\mathcal{E}(S)\ket\psi\ket- = (S\ket\psi)\ket-$, it
suffices by linearity to consider the two cases where $\ket\psi =
\ket0$ and $\ket\psi = \ket1$. When $\ket\psi = \ket0$ we have
\begin{align*}
  \mathcal{E}(S)\ket0\ket- & = CV \ket0\ket- = \ket0\ket- = (S\ket0)\ket-
\end{align*}
and when $\ket\psi = \ket1$ we have
\begin{align*}
  \mathcal{E}(S)\ket1\ket- 
  & = CV \ket1\ket- \\
  & = \ket1(V\ket-) \\
  & = \ket1(HSH\ket-) \\
  & = \ket1(HS\ket1) \\
  & = \ket1(H(i\ket1)) \\
  & = i\ket1(H\ket1) \\
  & = i\ket1\ket- \\
  & = (S\ket1)\ket-
\end{align*}
The correctness of the catalytic embedding
$\mathcal{E}(T)\ket\psi\ket{T} = (T \ket\psi)\ket{T}$ was established
in \cite[Thm.~4.1]{cyclotomic}.

\subsection{Resource state injection}
The circuit in Fig.~\ref{fig:minusinj} computes the unitary
$$
U = \frac{1}{2} \begin{pmatrix}
   i &  1 &  1 & -i \\
  -i &  1 &  1 &  i \\
   1 &  i & -i &  1 \\
   1 & -i &  i &  1
\end{pmatrix}
$$
so when executed with $\ket{00}$ as input yields
\begin{align*}
  U\ket{00}
  & = \tfrac{1}{2}\left(i\ket{00} - i\ket{01} + \ket{10} + \ket{11}\right) \\
  & = \tfrac{1}{\sqrt{2}}\left(i\ket0\ket- + \ket1\ket+\right)
\end{align*}
Hence, when the first qubit is measured to be in the state $\ket0$
(which happens with probability $\tfrac{1}{2}$) the second qubit is in
the state $i\ket-$.
\end{document}